\documentclass[useAMS,usenatbib,usegraphicx]{mn2e}
\usepackage{epsf}

\title[Detailed cluster lensing profiles]
{Detailed cluster lensing profiles at large radii and the impact on
cluster weak lensing studies}

\author[M. Oguri and T. Hamana]
{Masamune Oguri and Takashi Hamana\\
Division of Theoretical Astronomy, National Astronomical Observatory
of Japan, 2-21-1 Osawa, Mitaka, Tokyo 181-8588, Japan.} 

\begin{document}

\date{\today}

\voffset- .5in

\pagerange{\pageref{firstpage}--\pageref{lastpage}} \pubyear{}

\maketitle

\label{firstpage}

\begin{abstract}
Using a large set of ray-tracing in $N$-body simulations, we examine
lensing profiles around massive dark haloes in detail, with a
particular emphasis on the profile at around the virial radii. We
compare radial convergence profiles, which are measured accurately in
the ray-tracing simulations by stacking many dark haloes, with our
simple analytic model predictions. Our analytic models consist of a main
halo, which is modelled by the Navarro-Frenk-White (NFW) density
profile with three different forms of the truncation, plus the
correlated matter (2-halo term) around the main halo. We find that the
smoothly truncated NFW profile best reproduces the simulated lensing
profiles, out to more than 10 times the virial radius. We then use
this analytic model to investigate potential biases in cluster weak
lensing studies in which a single, untruncated NFW component is
usually assumed in interpreting observed signals.   We find that
cluster masses, inferred by fitting reduced tangential shear profiles
with the NFW profile, tend to be underestimated by $\sim 5-10\%$ if
fitting is performed out to $\sim 10'-30'$. In contrast, the
concentration parameter is overestimated typically by $\sim 20\%$ for
the same fitting range. We also investigate biases in computing the
signal-to-noise ratio of weak lensing mass peaks, finding them to be
$\la 4\%$ for significant mass peaks. In the Appendices, we provide
useful formulae for the smoothly truncated NFW profile.
\end{abstract}


\section{Introduction}
\label{sec:intro}

Cosmological applications of clusters of galaxies rely on the
inference of cluster masses. A challenge lies in the fact that 
clusters are dominated by dark matter, which accounts for 
$\sim 80\%$ of the mass of the universe. While the distribution of
dark matter in clusters can accurately be predicted using $N$-body
simulations, it is difficult to relate the dark matter distribution
with observable characteristics of clusters, such as gas temperatures,
optical/X-ray luminosities, and member galaxy distributions, because
it involves many complicated physical processes associated with gas
cooling/heating and star formations. Indeed, recent cosmological
constraints from X-ray analysis of clusters has mainly been limited by 
various astrophysical uncertainties \citep[e.g.,][]{vikhlinin09,mantz10}.

Weak gravitational lensing provides a direct probe of dark matter
distributions in clusters. This method makes use of coherent
tangential distortion of distant background galaxies induced by 
deep potential wells of clusters. For massive clusters, we can detect
weak lensing signals out to virial radii, which allows direct
measurements of virial masses
\citep[e.g.,][]{dahle06,broadhurst08,mahdavi08,okabe08,oguri09,umetsu09,umetsu11,medezinski10,okabe10}. Even
for less massive clusters, 
one can measure an average mass of a given cluster sample by adding
up lensing signals for clusters in a given sample. This stacked lensing
technique has been applied to the Sloan Digital Sky Survey data 
to obtain constraints on cosmological parameters
\citep{mandelbaum07,johnston07,sheldon09,rozo10}, and will offer a
promising mass-calibration method in future wide-field optical imaging
surveys \citep{oguri11,rozo11}. In addition, stacked lensing signals
much beyond virial radii probe clustering of massive haloes, which
contains complementary information on halo masses \citep{oguri11}.

Weak lensing also provides a fascinating way to identify massive haloes in
a way unbiased to the physical state of the baryon, by searching for
peaks in the mass map reconstructed from lensing shear measurements
\citep*{schneider96,vanwaerbeke00,hamana04,hennawi05,maturi05,pace07}.
This idea has been applied to real data to demonstrate that massive
clusters can indeed be identified 
\citep{miyazaki02,miyazaki07,gavazzi07,schirmer07,hamana09,kubo09}. Such
shear-selected cluster catalogs not only offer a unique opportunity to
study the relation between mass and light in a statistical manner
\citep[e.g.,][]{geller10}, but also may provide an alternative way to
constrain cosmological parameters through number counts of peaks 
\citep[e.g.,][]{kratochvil10}. 

The interpretation of weak lensing data is usually made by comparing
observed signals with analytical model predictions. For the analytic
calculations of lensing properties, it is customary to adopt a density
profile proposed by \citet[][hereafter NFW]{navarro97}.  The NFW
profile has widely been used to extract information on cluster masses
from weak lensing data. However, results of such analysis can be
biased if the assumption on the NFW profile is not accurate. For
instance, lensing signals are determined by all matter distributions
along the line-of-sight, which cause the scatter and bias in mass
estimates
\citep{hoekstra03,dodelson04,deputter05,marian10,mandelbaum10,becker11,hoekstra11}, 
which can modify lensing signals particularly at large radii. Indeed,
earlier work using $N$-body simulations has found the significant
contribution of correlated matter around haloes
\citep[e.g.,][]{mandelbaum05,hayashi08,tavio08,cacciato09,hilbert10,masaki11}. 
If the true density profile deviates from the NFW profile, it can also
cause a systematic bias. Understanding such bias is clearly important
for attempts to use clusters as a cosmological probe.

In this paper, we investigate cluster lensing profiles in details using
a large set of ray-tracing in $N$-body simulations. We pay particular
attention to lensing profiles around virial radii where the origin of
lensing signals should change from main haloes to the correlated matter
around clusters. Thanks to the large number of ray-tracing
realisations, we can study lensing profiles out to very large radii,
typically several tens time virial radii of clusters. We consider an
analytic model that better fit the simulated signals, which is then
used to explore potential biases of several weak lensing studies
originating from the assumption of the NFW profile. 

The structure of this paper is as follows. In
Section~\ref{sec:theory} we present analytic models adopted in the
paper. Detailed comparisons with ray-tracing simulations are made 
in Section~\ref{sec:rayt}. Using the analytic model calibrated by the
ray-tracing, we study potential biases in various lens studies
in Section~\ref{sec:bias}. We summarise the results in
Section~\ref{sec:summary}. In Appendices~\ref{sec:bmo} and
\ref{sec:fourier}, we provide formulae which should be useful for
various calculations with the smoothly truncated NFW profile. 

\section{Modelling Halo Mass Profiles}
\label{sec:theory}

\subsection{Main Halo}
\label{sec:theory:1h}

In this paper, we study the radial profiles of the convergence and
tangential shear around clusters. At small scales the profiles are
dominated by the signals from the dark haloes associated with the
clusters (the so-called 1-halo term). The current most popular model
of the dark halo density profile is the profile proposed by NFW, which
is defined by the following form (hereafter the NFW profile): 
\begin{equation}
\rho_{\rm NFW}(r)=\frac{\rho_s}{(r/r_s)(1+r/r_s)^2}.
\label{eq:nfw}
\end{equation}
The density parameter $\rho_s$ is related to the virial mass 
$M_{\rm vir}$ defined such that the average density within the virial
radius becomes equal to the nonlinear overdensity $\Delta_{\rm vir}$,
which we compute using the spherical collapse model \citep[see,
  e.g.,][]{nakamura97}, times the mean matter density of the
universe. Specifically, $\rho_s$ is described as 
\begin{equation}
\rho_s=\frac{\Delta_{\rm vir}(z)\bar{\rho}_m(z)c_{\rm vir}^3}
{3m_{\rm nfw}(c_{\rm vir})}=\frac{M_{\rm vir}}
{4\pi r_s^3m_{\rm nfw}(c_{\rm vir})},
\end{equation}
where $c_{\rm vir}$ is the so-called concentration parameter defined
by
\begin{equation}
c_{\rm vir}\equiv \frac{r_{\rm
    vir}}{r_s}=\frac{1}{r_s}\left[\frac{3M_{\rm vir}}
{4\pi\Delta_{\rm vir}(z)\bar{\rho}_m(z)}\right]^{1/3},
\end{equation}
and $m_{\rm nfw}(c_{\rm vir})$ defined by
\begin{equation}
m_{\rm nfw}(c_{\rm vir})\equiv\int_0^{c_{\rm vir}}
\frac{x}{(1+x)^2}dx=\ln(1+c_{\rm vir})-\frac{c_{\rm vir}}{1+c_{\rm vir}}.
\end{equation}
The concentration parameter is known to be correlated with the halo
mass and redshift. When necessary, we adopt the following relation:
\begin{equation}
c_{\rm vir}(M_{\rm vir},z)=7.26\left(\frac{M_{\rm
    vir}}{10^{12}h^{-1}M_\odot}\right)^{-0.086}\left(1+z\right)^{-0.71},
\label{eq:concentation}
\end{equation}
which was derived from $N$-body simulations assuming best-fit
cosmological parameters in the Wilkinson Microwave Anisotropy Probe
(WMAP) third year results \citep{maccio08}, with the additional
redshift dependence based on the simulation result of \citet{duffy08}.
A well-known advantage of the NFW profile in lensing studies is that
there are analytic expressions for the radial profiles of the
deflection angle, convergence, and shear
\citep{bartelmann96,wright00}. 

However, the NFW profile is not well-defined in the sense that 
the enclosed mass diverges logarithmically. Thus
\citet{takada03a,takada03b} considered lensing by the NFW profile
truncated at the virial radius (hereafter the TJ profile): 
\begin{equation}
\rho_{\rm TJ}(r)=\frac{\rho_s}{(r/r_s)(1+r/r_s)^2}\Theta(r_{\rm vir}-r),
\label{eq:tj}
\end{equation}
with $\Theta(x)$ being the Heaviside step function. The lensing
properties of this profile can also be computed analytically
\citep[see][]{takada03a,takada03b}.

One potential problem of the TJ profile is that the shear and
convergence profiles are not differentiable at the truncation radius,
which causes the divergence in the flexion profile. 
\citet*{baltz09} proposed a different form of the truncation 
(hereafter the BMO profile):
\begin{equation}
\rho_{\rm BMO}(r)=\frac{\rho_s}{(r/r_s)(1+r/r_s)^2}
\left(\frac{r_t^2}{r^2+r_t^2}\right)^n,
\label{eq:bmo}
\end{equation}
where $r_t$ is the truncation radius. In the paper we mainly use the
following dimensionless truncation radius
\begin{equation}
\tau_v\equiv \frac{r_t}{r_{\rm vir}}.
\label{eq:tauv}
\end{equation}
While both $n=1$ and $2$ have
been considered in \citet{baltz09}, in this paper we consider only
$n=2$ for simplicity. In the case, at large radii the density profile
behaves as  $\rho_{\rm BMO}(r)\propto r^{-7}$, and therefore the
enclosed mass converges quickly. The three-dimensional enclosed mass 
of the profile is 
\begin{equation}
M_{\rm bmo}(r)=4\pi\rho_s r_s^3 m_{\rm bmo}(x=r/r_s),
\label{eq:mbmo_dim}
\end{equation}
\begin{eqnarray}
m_{\rm
  bmo}(x)&=&\frac{\tau^2}{2(\tau^2+1)^3(1+x)(\tau^2+x^2)}\nonumber\\
&&\hspace*{-16mm}\times\Bigl[(\tau^2+1)x\left\{x(x+1)-\tau^2(x-1)(2+3x)
-2\tau^4\right\}\nonumber\\
&&\hspace*{-16mm}+\tau(x+1)(\tau^2+x^2)\left\{2(3\tau^2-1)
{\rm arctan}(x/\tau)\right.\nonumber\\
&&\hspace*{-16mm}\left.+\tau(\tau^2-3)\ln(\tau^2(1+x)^2/(\tau^2+x^2))\right\} \Bigr],
\label{eq:mbmo_nodim}
\end{eqnarray}
where $\tau\equiv r_t/r_s = \tau_vc_{\rm vir}$ and we have fixed $n=2$
in equation.~(\ref{eq:bmo}). Again, the advantage
of the choice of this specific form of the truncation is that lensing
properties can be computed analytically \citep[see][]{baltz09}.  

In equation~(\ref{eq:bmo}), we adopted the same density parameter
$\rho_s$ as used in the NFW profile, with $M_{\rm vir}$ as an input
parameter of this model. This virial mass $M_{\rm vir}$,
strictly speaking, is different from a true virial mass defined from
the requirement of the nonlinear overdensity. However, one advantage
of this choice of the normalisation is that the ``virial mass'' of
this model, $M_{\rm vir}$, becomes rather close to the total mass 
$M_{\rm tot}$. From the definition above, we can show that these two
masses are related by (again for $n=2$) 
\begin{equation}
M_{\rm tot}=\frac{m_{\rm tot}}{m_{\rm nfw}(c_{\rm vir})}M_{\rm vir},
\label{eq:mbmo_dim_tot}
\end{equation}
\begin{eqnarray}
m_{\rm tot}&\equiv & m_{\rm bmo}(\infty)\hspace*{2mm}=\hspace*{2mm}
\frac{\tau^2}{2(\tau^2+1)^3}\nonumber\\
&&\hspace*{-12mm}\times\left[(3\tau^2-1)(\pi\tau-\tau^2-1)+2\tau^2(\tau^2-3)\ln\tau\right].
\label{eq:mbmo_nodim_tot}
\end{eqnarray}
Typically we have $M_{\rm tot}/M_{\rm vir}\sim 1-1.3$ and $M(<r_{\rm
  vir})/M_{\rm vir}\sim 0.9$ in the parameter range we are interested in.

\subsection{Correlated matter}
\label{sec:theory:2h}

At the large scale, the correlated matter around the cluster (the
so-called 2-halo term) contributes to the lensing profile
\citep[e.g.,][]{johnston07}. Using the
Limber's approximation, we compute the 2-halo term contribution
for a cluster with mass $M$ and at redshift $z$ as 
\citep[e.g.,][]{oguri11}
\begin{equation}
\kappa_{\rm 2h}(\theta)=\int\!\frac{\ell d\ell}{2\pi}
J_0(\ell\theta) \frac{\bar{\rho}_{m}(z)b_h(M)}{(1+z)^3\Sigma_{\rm crit}D_A^2(z)}
P_m\left(k_\ell; z\right),
\label{eq:kappa2h}
\end{equation}
where $k_\ell\equiv \ell/\{(1+z)D_A(z)\}$, $D_A(z)$ is the angular
diameter distance, $\Sigma_{\rm crit}$ is the (physical) critical mass
density for lensing, $P_m(k)$ is the linear matter power spectrum, and
$b_h(M)$ is the halo bias. We can compute the corresponding radial
profile of the tangential shear $\gamma_{\rm T}(\theta)$ simply by
replacing the zero-th order Bessel function $J_0(x)$ in
equation~(\ref{eq:kappa2h}) to the second order Bessel function
$J_2(x)$. The power spectrum $P_m(k)$ is computed using the
approximation of the transfer function presented by
\citet{eisenstein98}.  

We adopt a model of the halo bias $b_h(M)$ by \citet{tinker10}, which
has been calibrated using a large set of $N$-body simulations. The
model has been presented as a function of several different mean
overdensities $\Delta$. We adopt a model with $\Delta=800$, a larger
overdensity than the virial overdensity of $\Delta\approx 300$, so
that the details of the truncation near the virial radius do not
affect the conversion of the masses very much. For a given set of
$M_{\rm vir}$ and $c_{\rm vir}$, we compute the mass for $\Delta=800$
assuming the NFW profile \citep[see, e.g.,][]{hu03}.

\section{Comparison with ray-tracing simulations}
\label{sec:rayt}

\subsection{Simulated data}

In order to study the detailed radial profiles of lensing signals
around massive haloes, we employ a large set of ray-tracing
simulations presented by \citet{sato09}. The ray-tracing simulations
are based on $200\times2$ realisations of $N$-body simulations with
the box sizes of 240 (particle mass $m_p=5.4\times 10^{10}h^{-1}M_\odot$)
and $480h^{-1}$Mpc ($m_p=4.4\times 10^{11}h^{-1}M_\odot$). The number of
particles in each $N$-body run is 
$256^3$. Outputs of the small and large box simulations at different
redshifts are then placed to construct a light cone of the
$5^\circ\times5^\circ$ region. We use the standard multiple lens plane
algorithm to simulate lensing by intervening matter. We trace $2048^2$
light rays backward from the observer, resulting in the angular grid
size of $0.15'$. Thus our ray-tracing simulations take account of all
density fluctuations along the line-of-sight, not just the matter
distributions around massive haloes.  While lens planes have been
constructed out to $z=3$, in this paper we consider only one source
redshift of $z_s=0.997$, which is a typical mean source redshift of
weak lensing analysis. Thanks to the large number of $N$-body runs, 1000 
independent ray-tracing realisations can finally be generated by
randomly shifting the simulation boxes. The total effective area of  
25000deg$^2$ thus allows us to compute the average radial profiles
quite accurately. Interested readers are referred to
\citet{sato09} for more details of $N$-body simulations and
ray-tracing technique.  The assumed cosmological parameters are 
the matter density $\Omega_M=0.238$, baryon density $\Omega_b=0.042$, 
cosmological constant $\Omega_\Lambda=0.762$, spectral index
$n_s=0.958$, the normalisation of the matter fluctuation
$\sigma_8=0.76$, and the Hubble parameter $h=0.732$, which we adopt in
our analytic calculations too.  

In the $N$-body simulations we identify dark haloes using the
friend-of-friend (FOF) algorithm with the linking length of $b=0.2$ to
derive the total mass $M_{\rm FOF}$ of each halo. We compute average
convergence and tangential shear profiles in each mass and redshift
bin. We consider 5 mass bins centred at $M_{\rm FOF}=10^{13.75}$,
$10^{14}$, $10^{14.25}$,  $10^{14.5}$, and $10^{14.75}h^{-1}M_\odot$,
with the bin size of $\Delta(\log M_{\rm FOF})=0.1$, and 7
redshift bins centred at $z=0.1$, 0.2, 0.3, 0.4, 0.5, 0.6, and 0.7, 
with the bin size of $\Delta z=0.04$. The number of haloes used in
each bin ranges from $\sim 20$ to $\sim 20000$. When stacking the
profiles, we carefully choose a centre of each halo by identifying a
peak in the convergence map in order to avoid any off-centring effect
which essentially smears out central profiles \citep[see][]{oguri11}.
The average profiles are computed as a function of 
$\theta/\theta_{\rm vir}$, where $\theta_{\rm vir}$ is the virial
radius computed assuming $M_{\rm vir}=M_{\rm FOF}$. We derive profiles
in the radius range of 
$10^{-2}<\theta/\theta_{\rm vir}<50$, with the bin size of
$\log(\theta/\theta_{\rm vir})=0.1$.

\begin{figure}
\begin{center}
 \includegraphics[width=0.9\hsize]{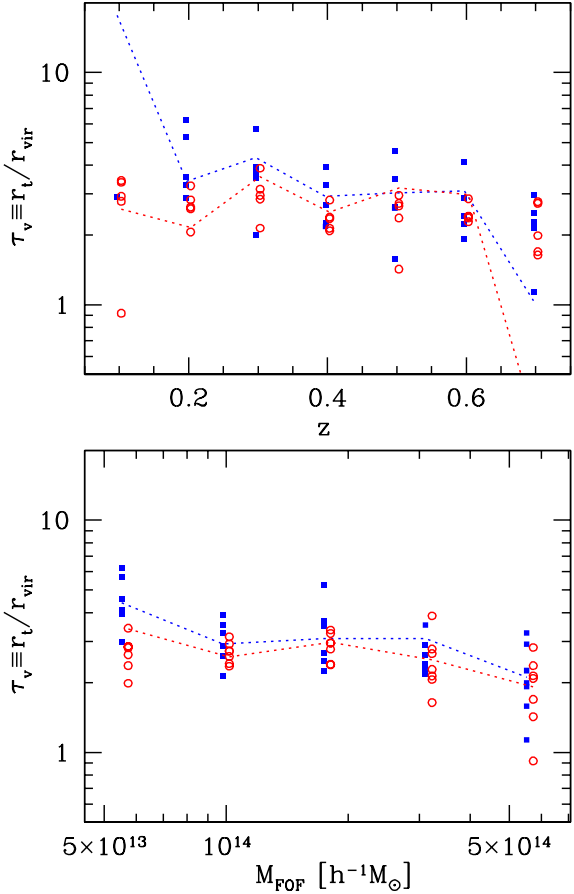}
\end{center}
\caption{Bets-fit values of the dimensionless truncation radius
$\tau_v$  (eq.~[\ref{eq:tauv}]) in the BMO profile, for different
redshift ({\it upper}) and mass ({\it lower}) bins. Filled squares
plot the case when the concentration parameter is computed from
equation~(\ref{eq:concentation}), whereas open circles are the case
when the concentration parameter is fitted simultaneously. Dotted
lines connect median values of $\tau_v$ in different redshift and mass
bins. The median values of all bins are $\tau_v=3.0$ for the fixed
concentration parameter, and $\tau_v=2.6$ for the fitted concentration
parameter. 
\label{fig:tau}}
\end{figure}

\subsection{Fitting method}

We fit the average convergence profile in each bin using the models
described in Sec.~\ref{sec:theory}. We basically fit the average
profiles in the simulations with the sum of the 1-halo and 2-halo
components, $\kappa(\theta)=\kappa_{\rm 1h}(\theta)+\kappa_{\rm
  2h}(\theta)$, with three different models of the 1-halo components
as described in Sec.~\ref{sec:theory:2h}. The standard $\chi^2$ method is
employed for fitting, with the error of convergence in each radial bin
derived from the scatter in the simulations. While the average
profiles are derived for given ranges of $M_{\rm FOF}$, it is not
clear whether $M_{\rm FOF}$ coincides with the virial mass 
$M_{\rm vir}$ used in the analytic models. In fact, previous work
using $N$-body simulations has found that there are systematic bias
and considerable scatter between $M_{\rm FOF}$ and $M_{\rm vir}$
\citep[e.g.,][]{white01,tinker08}.  Therefore, we perform fitting
with leaving the virial mass $M_{\rm vir}$ as a free parameter to
take account of such systematic effect. Unless otherwise stated, we
regard the concentration parameter $c_{\rm vir}$ as a free parameter,
although our main conclusions are unchanged even if we fix the
concentration parameter to the value obtained from
equation~(\ref{eq:concentation}). Thus there are 2 fitting parameters
($M_{\rm vir}$, $c_{\rm vir}$) for  the NFW and TJ profiles, and 3
fitting parameters ($M_{\rm vir}$, $c_{\rm vir}$, $\tau_v$) for the
BMO profile.    

It has been known that very central density profiles of dark haloes in
$N$-body simulations are not reliable because of several numerical
effects such as the two-body relaxation and the finite time step size.
Based on the detailed analysis of \citet{fukushige01}, we estimate
that density profiles are reliable down to  $\sim 0.08 r_{\rm vir}$
for our $N$-body simulations in which $N\sim 10^4$ particles are
included in each halo analysed in the present paper.  Indeed, our
analysis results also indicate that convergence and tangential shear
profiles near the very centre ($\theta/\theta_{\rm vir}\sim 0.01$)
tend to be significantly smaller than our analytic model calculations
(see below). Thus, in all mass and redshift bins fitting is performed
in the radial bins $\theta/\theta_{\rm vir}\geq 0.079$ to make sure
that our fitting results are not affected by numerical artifacts.

\subsection{Result}

\begin{figure*}
\begin{center}
 \includegraphics[width=0.85\hsize]{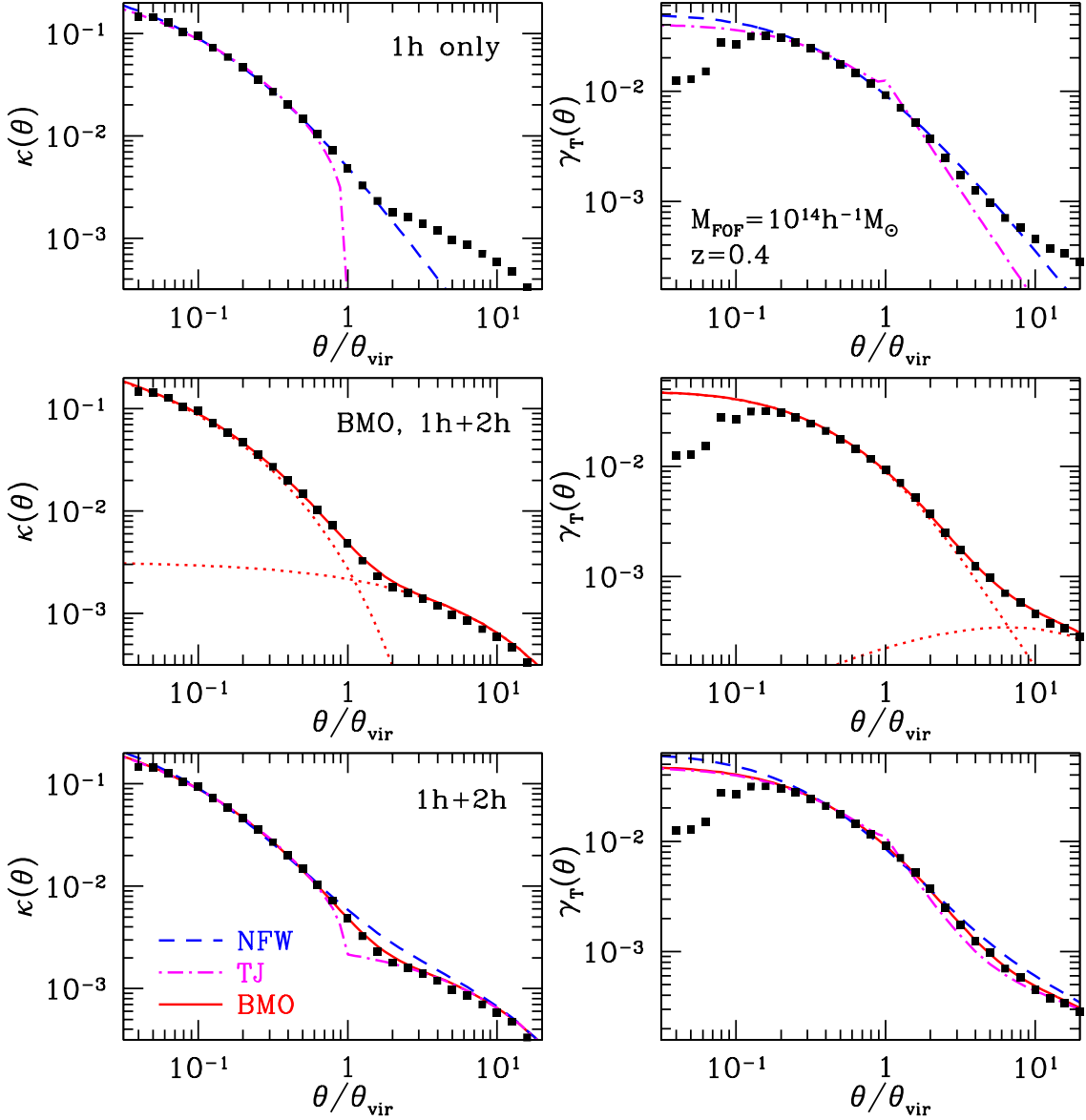}
\end{center}
\caption{Example of fitting results for the mass bin $M_{\rm
    FOF}=10^{14}h^{-1}M_\odot$ and the redshift bin $z=0.4$. Left
  panels show convergence profiles, whereas right panels display
  tangential shear profiles. Curves are best-fit results for three
  different main halo models, the NFW profile ({\it dashed}), the TJ
  profile ({\it dash-dotted}), and the BMO profile ({\it solid}). 
  The average profiles from ray-tracing simulations are indicated by
  filled squares. {\it Top:} Results when only 1-halo term (i.e., no
  2-halo term) is considered. The result for the BMO is similar to the
  NFW result, and therefore not shown. {\it Middle:} Results
  for the BMO profile. Contributions from 1-halo and 2-halo terms
  are shown by dotted lines. {\it Bottom:} Results for all the 3
  profiles are compared. 
\label{fig:ex_prof}}
\end{figure*}

First we check the best-fit values of the truncation radius $\tau_v$ 
(eq.~[\ref{eq:tauv}]) in the BMO profile. Figure~\ref{fig:tau} show
the best-fit values for different redshift and mass bins. We find the
best-fit values to be $\sim 2-3$, showing no strong dependence on the
mass and redshift. We consider two cases, the case that the
concentration parameter is fixed to the value computed in
equation~(\ref{eq:concentation}) and the case that the concentration
parameter is also fitted to the data. We find that the results of both
cases agree reasonably well with each other, which implies that the
best-fit values of $\tau_v$ is not sensitive to how the concentration
parameter is treated. The median values are $\tau_v=3.0$ when the
concentration parameter is fixed, and $\tau_v=2.6$ when the
concentration parameter is fitted to the data.

To illustrate how well the different main halo models can reproduce
the average profiles in ray-tracing simulations, in
Figure~\ref{fig:ex_prof} we show the comparisons of convergence and
tangential shear profiles for a representative case. As clearly shown
in the Figure, the BMO profile reproduce profiles in the ray-tracing
simulations quite successfully for a wide range in radii. In contrast, 
the NFW and TJ profiles are less successful in fitting the profile
around the virial radius, where the transition between 1-halo and
2-halo terms occurs. The NFW profile tends to overpredict the
convergence profile, whereas the TJ profile clearly underpredicts the
profile. For comparison, we fit the convergence profile using the
1-halo term only for the NFW and TJ profile. We find that the NFW
profile can fit the total average profile quite well out to
$\theta/\theta_{\rm vir}\sim 2$, beyond which the NFW profile clearly
underpredicts the profile. The situation is worse for the TJ profile
for which we can see the significant discrepancy already at
$\theta/\theta_{\rm vir}\sim 1$. 

The Figure also indicates that the analytic and simulated convergence
profiles agree well with each other down to very small radii of
$\theta/\theta_{\rm vir}\sim 0.05$, whereas the tangential shear
profiles show large deviations already at $\theta/\theta_{\rm vir}\sim
0.1$. One reason for this is that shear signals are non-local. Since
the tangential shear signal at certain radius reflects all the mass
distributions at smaller radii, numerical effects appear at larger
radii in the tangential shear profiles than in the convergence
profiles. 

\begin{figure}
\begin{center}
 \includegraphics[width=0.9\hsize]{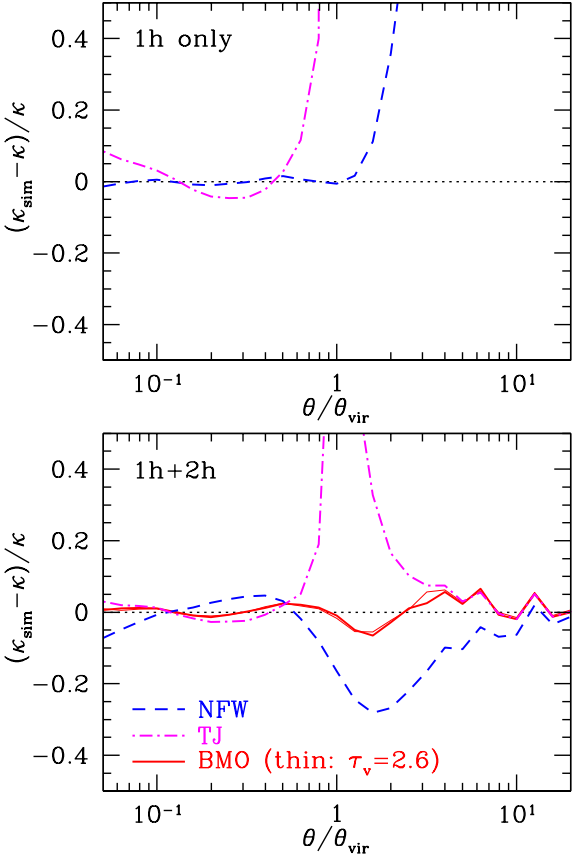}
\end{center}
\caption{Residuals of convergence profile fitting (see 
  Figure~\ref{fig:ex_prof} for a representative example) averaged over
  all redshift and mass bins, plotted as a function of the normalised
  radius $\theta/\theta_{\rm vir}$. Specifically we define the
  residual as ($\kappa_{\rm sim}-\kappa)/\kappa$, where $\kappa_{\rm
    sim}$ is the average convergence profile from ray-tracing
  simulations and $\kappa$ is the convergence profile of the best-fit
  analytic model. Lines are same as Figure~\ref{fig:ex_prof}.
  {\it Upper:} Residuals for fitting when only 1-halo term is
  considered (see also top panels of Figure~\ref{fig:ex_prof}). 
  Again, the result for the BMO is similar to the NFW result.  
  {\it Lower:} Residuals when both 1-halo and 2-halo terms are
  included in the analytic model (see also bottom panels of
  Figure~\ref{fig:ex_prof}). The thin solid line indicates the result
  when the truncation radius in the BMO profile is fixed to
  $\tau_v=2.6$, the median value among fitting results for all
  redshift and mass bins (see Figure~\ref{fig:tau}).
  \label{fig:residual}}
\end{figure}

Figure~\ref{fig:residual} show residuals of the convergence profile,
i.e., the fractional difference between profiles in simulations and
best-fit analytic models, averaged over all redshift and mass bins.
We find that the BMO profile can fit the
convergence profiles in simulations quite well ($<5\%$) for a wide
range of radii from the core of main haloes to more than 10 times the
virial radii. We find the NFW profile generally overpredicts the
convergence profile by $\sim 20-30\%$ at $\theta/\theta_{\rm vir}\sim
2$. On the other hand, the TJ profile grossly underpredicts the
convergence profile at $\theta/\theta_{\rm vir}\sim 1$. Thus we
confirm the result shown in Figure~\ref{fig:ex_prof} that the BMO
profile can best describe convergence profiles in ray-tracing
simulations. Again, we check residuals for the case that we include
only 1-halo term in the analytic model, and find that the NFW profile
can fit the convergence profiles quite well out to $\theta/\theta_{\rm
  vir}\sim 1$, but begins to underpredicts the profile quickly beyond
the radius.

\section{Impact of inaccurate profiles on cluster weak 
lensing studies} 
\label{sec:bias}

In most weak lensing studies, the NFW profile has been adopted in
comparing with observed lensing signals, without including any
contribution from the 2-halo term. However, the difference between
assumed and true profiles can induce systematic biases in the 
interpretation of results. In this section, we investigate such
systematic biases, assuming the analytic model calibrated by
ray-tracing simulations (see Sec.~\ref{sec:rayt}) as a true cluster
lensing profile. 

\subsection{Shear profile fitting}

The most popular method to measure cluster masses from weak lensing
data is to fit the tangential shear profile with analytic model
predictions, for which the untruncated NFW profile is usually adopted
as the analytic model. Here we estimate how cluster masses derived by
such shear profile fitting can be biased due to the deviation of the
true cluster lensing profile from the one computed from the NFW
profile.   

\begin{figure}
\begin{center}
 \includegraphics[width=0.9\hsize]{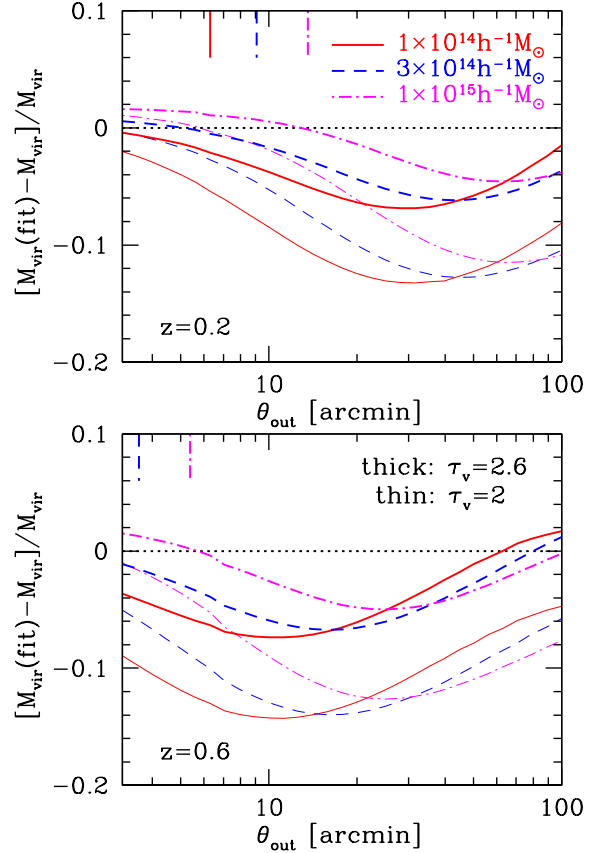}
\end{center}
\caption{Biases in weak lensing mass estimates from shear profile
  fitting, as a function of the outermost fitting radius $\theta_{\rm
    out}$, derived by adopting the BMO profile as the true lensing
  profile. The bias is defined as $\left[M_{\rm vir}({\rm fit})-M_{\rm
  vir}\right]/M_{\rm vir}$, where $M_{\rm vir}$ is the input virial
  mass and $M_{\rm vir}({\rm fit})$ is the best-fit virial mass to the
  input reduced shear profile, using the NFW profile with no 2-halo
  term as a model for fitting. Note that the innermost fitting radius
  is fixed to 
  $\theta_{\rm in}=1'$. We consider three input halo masses, $M_{\rm
    vir}=10^{14}h^{-1}M_\odot$ ({\it solid}),
  $3\times10^{14}h^{-1}M_\odot$ ({\it dashed}), and
  $10^{15}h^{-1}M_\odot$ ({\it dash-dotted}). The halo redshifts are
  $z=0.2$ ({\it upper}) and $0.6$ ({\it lower}). Thick and thin lines
  indicate results for the input truncation radii of $\tau_v=2.6$ and
  $\tau_v=2$, respectively. Vertical lines show virial radii for these
  cluster masses.
  \label{fig:bias}}
\end{figure}

\begin{figure}
\begin{center}
 \includegraphics[width=0.9\hsize]{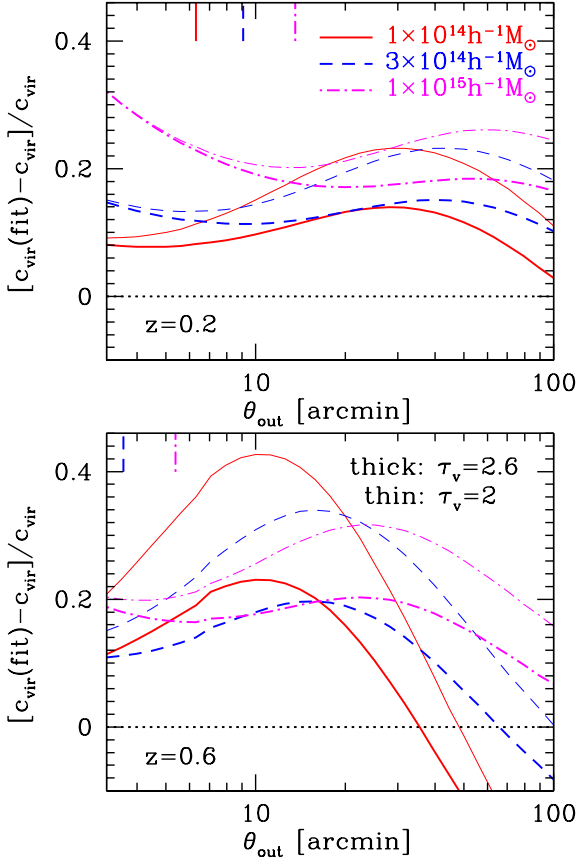}
\end{center}
\caption{Same as Figure~\ref{fig:bias}, but biases in
  estimates of the concentration parameter $c_{\rm vir}$ are shown. 
  \label{fig:bias_c}}
\end{figure}

In actual weak lensing analysis of observed data, we measure reduced
shear profiles around clusters, which are defined by $g(\theta)\equiv
\gamma_{\rm T}(\theta)/\left[1-\kappa(\theta)\right]$. For a given
reduced shear profile, we usually derive the best-fit mass by
minimizing the following $\chi^2$:
\begin{equation}
\chi^2=\sum_i\frac{\left[ g(\theta_i)-
g_{\rm NFW}(\theta_i; M_{\rm vir},c_{\rm vir})\right]^2}{\sigma_i^2},
\label{eq:chi2}
\end{equation}
where $g(\theta_i)$ is observed reduced shear profile at the radius
$\theta=\theta_i$ and $g_{\rm NFW}(\theta_i; M_{\rm vir},c_{\rm
  vir})$ is the corresponding analytic model prediction assuming the
NFW profile with no 2-halo term. We compute $g(\theta_i)$ assuming the
BMO profile which has been shown to best reproduce lensing profiles
in ray-tracing simulations (Sec.~\ref{sec:rayt}). Assuming the uniform
source galaxy density, we adopt the error $\sigma_i\propto
1/\theta_i$. The $\chi^2$ is computed in the range $\theta_{\rm
  in}<\theta<\theta_{\rm out}$ with an interval of $\Delta(\log
\theta)=0.05$. We fix the innermost radius to $\theta_{\rm in}=1'$,
and see how the best-fit mass $M_{\rm vir}$ differs from the input
mass as a function of $\theta_{\rm out}$. In fitting, we vary both
$M_{\rm vir}$ and $c_{\rm  vir}$ to search for the best-fit parameter
values. 

Figure~\ref{fig:bias} shows the results, the biases as a function of
$\theta_{\rm out}$. We find that adopting a single NFW profile tends
to underestimate virial masses for $\theta_{\rm out}\sim 10'-30'$, a
typical outermost radii adopted in the real analysis of cluster weak
lensing data. The exact amount of the biases appears to be sensitive
to $\tau_v$. For our canonical value of $\tau_v=2.6$, which was
the median value among fitting results for many redshift and mass bins
in ray-tracing simulations (see Figure~\ref{fig:tau}) virial masses
are underestimated by $\sim 5\%$. On the other hand, if we adopt a
slightly smaller value of $\tau_v=2$, virial masses can be underestimated
by more than 10\%. If we choose very large values of 
$\theta_{\rm out}$, the bias becomes smaller because of counteracting
effects of the overpredciton of signals near the virial radius and the
underprediction of the signals at very large radii where the 2-halo
term contributions dominates. Our result suggests that we can reduce
such mass estimation bias by restricting the fitting range small,
$\theta_{\rm out}\la \theta_{\rm vir}$. 

Recently, \citet{becker11} studied biases in weak lensing cluster
mass estimates in details using a cosmological $N$-body simulations,
and found that weak lensing masses are biased low by $\approx 5-10\%$. 
They adopted $\theta_{\rm out}=15'$, $20'$, $25'$, and found a larger
amount of the bias with increasing $\theta_{\rm out}$. Thus our 
results are consistent with their results, both qualitatively and
quantitatively, and moreover provide a physical explanation for their
finding. 

In Figure~\ref{fig:bias_c}, we show biases in concentration
parameter estimates. We find that the concentration parameter is
generally overestimated, typically by $\sim 20\%$. Clearly this is
related to the mass estimation bias studied above. In lensing analysis 
with the NFW profile, there is a well-known degeneracy between mass
and concentration parameter, and hence the underestimate of the mass
has to be compensated by the increase of the concentration parameter
in order to recover lensing signals near the cluster centre.

A caveat here is that the input $M_{\rm vir}$ in the BMO profile
differs from the true virial mass of the profile, which may make the
interpretation of this result somewhat difficult. In fact we can avoid
this problem by considering larger nonlinear overdensity $\Delta$,
such as $\Delta\approx 1600$ used by \citet{becker11}, because the
enclosed masses for the NFW and BMO profiles adopting the same $\rho_s$
should become more similar for larger values of $\Delta$. We confirm
that our results shown in Figures~\ref{fig:bias} and \ref{fig:bias_c}
similarly hold for different nonlinear overdensities.

\subsection{Weak lensing mass peak}

Next we discuss how the uncertainty of lensing profiles can affect the
interpretation and statistics of weak lensing mass peaks. 
The mass peak is usually characterised by the signal-to-noise ratio:
\begin{equation}
\nu=\frac{\kappa_{\rm peak}}{\sigma_{\rm noise}},
\label{eq:nu}
\end{equation}
where $\kappa_{\rm peak}$ is the filtered peak amplitude computed by
\begin{equation}
\kappa_{\rm peak}=\int \gamma_{\rm T}(\theta)Q(\theta) 2\pi\theta d\theta,
\end{equation}
with $Q(\theta)$ being the filter function. The noise $\sigma_{\rm
  noise}$ is related to the intrinsic ellipticity and the number
 density of source galaxies, for which we assume $\sigma_e=0.35$ and
  $n_{\rm gal}=30$~arcmin$^{-2}$. In this paper, we consider two
  filter functions. One is the Gaussian filter \citep{kaiser93}, 
\begin{equation}
Q(\theta)=\frac{1}{\pi\theta_{\rm in}^2}\frac{1-(1+x^2)e^{-x^2}}{x^2},
\label{eq:fil_ga}
\end{equation}
and the other is a filter proposed by \citet[][ hereafter HS05]{hennawi05},
\begin{eqnarray}
Q(\theta)&=&\frac{1}{\pi\theta_{\rm
    in}^2}\left[\frac{2\log(1+x)}{x^2}-\frac{2}{x(1+x)}-\frac{1}{(1+x)^2}\right]\nonumber\\
&&\times e^{-x^2/2x_{\rm out}^2},
\label{eq:fil_hs}
\end{eqnarray}
where $x\equiv \theta/\theta_{\rm in}$ and $x_{\rm out}\equiv
\theta_{\rm out}/\theta_{\rm in}$. We adopt $\theta_{\rm in}=1'$ for
the Gaussian filter, and $\theta_{\rm in}=0.5'$ and $\theta_{\rm
  out}=10'$ for the HS05 filter, which are typical values adopted in
the literature.

\begin{figure}
\begin{center}
 \includegraphics[width=0.9\hsize]{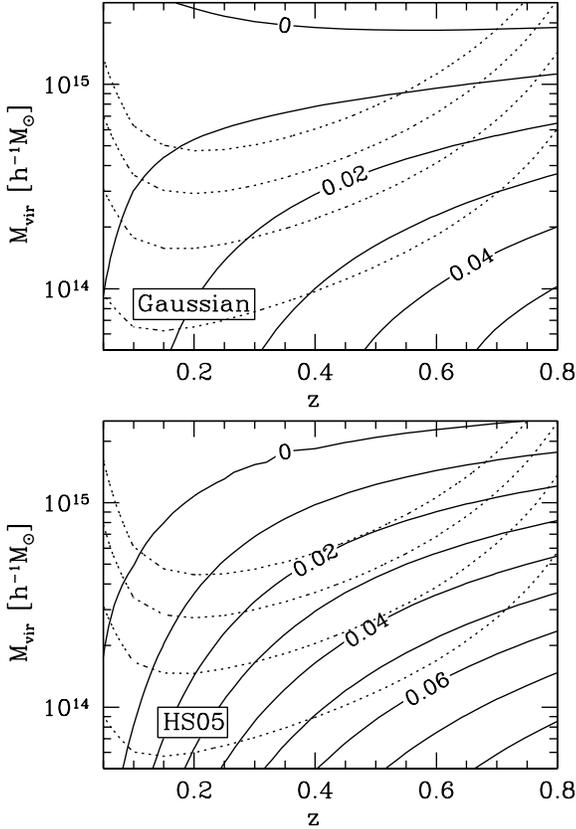}
\end{center}
\caption{Contours of the biases in the signal-to-noise ratio $\nu$
  defined in equation~(\ref{eq:nu}) in the mass-redshift plane. The
  bias is defined by $(\nu_{\rm NFW}-\nu)/\nu$, where $\nu_{\rm NFW}$
  is the signal-to-noise ratio computed using the NFW profile with no
  2-halo term, and $\nu$ is the true signal-to-noise ratio computed
  from the BMO profile with $\tau_v=2$. Note that the bias is even
  smaller for the canonical truncation radius of $\tau_v=2.6$. Dotted
  lines show contours for $\nu=3$, $5$, $7$, and $9$, from bottom to
  top. The top panel is the result for the Gaussian filter
  (eq.~[\ref{eq:fil_ga}]), whereas the bottom panel is for the HS05
  filter (eq.~[\ref{eq:fil_hs}]).  
 \label{fig:peak}}
\end{figure}

\begin{figure}
\begin{center}
 \includegraphics[width=0.9\hsize]{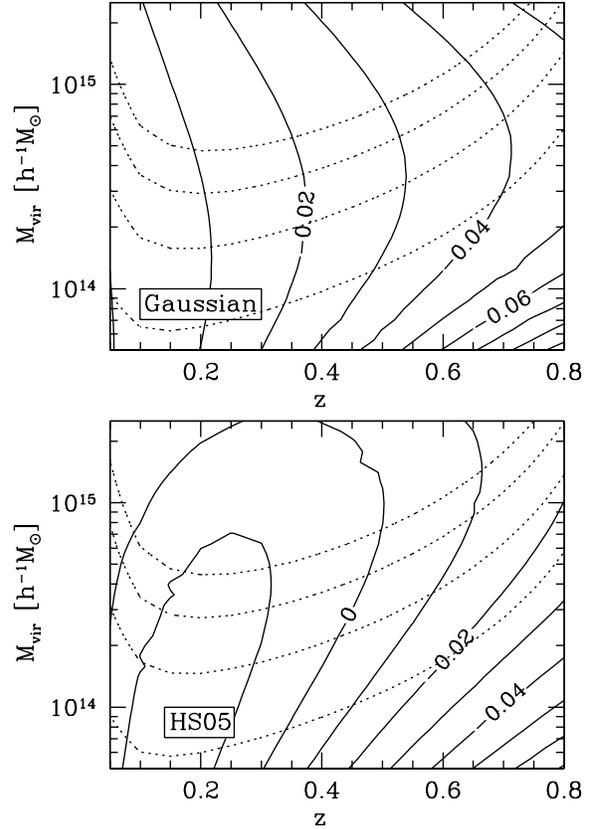}
\end{center}
\caption{Same as Figure~\ref{fig:peak}, but the biases for the TJ
  profile are shown.
  \label{fig:peak_tj}}
\end{figure}

We study the bias in $\nu$ assuming the true profile to be the BMO
profile. Figure~\ref{fig:peak} shows how values of $\nu$ are biased by
adopting the NFW profile with no 2-halo term. We find that the bias is
$\la 4\%$ for peaks with $\nu\sim 3$ and smaller for higher
$\nu$. The bias is not so large because $\nu$ is mainly determined by
the lensing profiles well inside the virial radius where the
difference of the profiles are rather small. Note that the result
shown in Figure~\ref{fig:peak} is for the truncation radius of
$\tau_v=2$, which is smaller than our canonical value of
$\tau_v=2.6$. The bias is even smaller if we adopt $\tau_v=2.6$.

In some previous work, the TJ profile has been used to compute $\nu$
instead of the NFW profile \citep[e.g.,][]{hamana04}. In
Figure~\ref{fig:peak_tj} we show the bias for the TJ profile. We
confirm that the bias has a different sign from the case for the NFW
profile, with a similar size of $\la 4\%$ for most of the parameter
range we are interested in. 

Therefore, in either case the effect of the outer profile on the
calculation of the peak height $\nu$ is not very large, although 
it can be important for high-precision analytic predictions.
We note that the result presented here is an averaged effect. In
practice, there is a considerable scatter on the amount of  
correlated matter around clusters, which can affect the statistics of
weak lensing mass peaks in various ways. The detailed study of
statistical properties of weak lensing mass peaks is currently
underway, which will be reported in a forthcoming paper (Hamana \&
Oguri, in preparation). 

\section{Summary and discussions}
\label{sec:summary}

We have studied detailed lensing profiles and their impact on various
cluster lensing studies. First, we compare our simple analytic model
predictions, which consist of three different 1-halo terms and a
2-halo term, with a large set of ray-tracing in $N$-body simulations,
with a particular emphasis on lensing profiles at around virial
radii of massive haloes. Our ray-tracing solves the full light
propagation from the source plane to the observer, and thus take
accounts of all the matter fluctuations along the line-of-sight. 
We find that the BMO profile, which is essentially a smoothly
truncated NFW profile, with the truncation radius of $\tau_v\sim 2-3$
with additional contribution from the 2-halo term best reproduces the
lensing profiles in ray-tracing simulations. In contrast, the
untruncated NFW profile tends to overpredict the convergence profiles
by $20-30\%$ at twice the virial radii when the 2-halo term is
added. The TJ (hard-truncation) profile largely underpredicts the
lensing signals at around the virial radii. 

Next, assuming the BMO profile with the 2-halo term as the ``true''
cluster lensing profile, we have investigated biases in cluster weak
lens studies coming from the adoption of the NFW profile as an
analytic model. We find that, if observed reduced shear profiles are
fitted by a single NFW component, as has commonly been done in cluster
weak lensing analysis, we can underestimate cluster masses up to
$5-10\%$. The bias is most significant when the outer boundary for
fitting is set to $\sim 10'-30'$ depending on the mass and redshift of
the cluster, or equivalently several times the virial radius. We can
reduce the bias by restricting the fitting range small with the outer 
boundary equal or smaller than the virial radius. On the other hand,
the concentration parameter tends to be biased high (up to $\sim 20\%$
or even higher) to compensate the underestimate of the mass. These
results based on the analytic model appear to be consistent with
recent numerical results by \citet{becker11}.

In the reduced shear fitting above, we have considered only 
intrinsic shapes of galaxies as a source of the error, i.e., ignored the
contribution of the large-scale structure (cosmic shear) to the error 
\citep[e.g.,][]{hoekstra03,dodelson04,hoekstra11}. The error from the
large-scale structure becomes increasingly important at larger radii,
and thus decreases the relative weight of bins at large radii,
implying that it reduces the bias in the mass and concentration
parameter estimates. The exact amount of the reduction of the bias,
however, depends on the source galaxy number density assumed. In real
cluster weak lensing analysis, colour cuts have often been applied in
order to minimise the dilution effect by cluster member galaxies,
which significantly decreases source galaxy number densities used for
the analysis. Because of this, the contribution of the large-scale
structure to the total error budget tends to be subdominant in real
data analysis \citep[e.g.,][]{oguri10}, but the large-scale structure
should still contribute significantly at very large radii, e.g.,
$\theta\ga 20'$.  

We have also examined the effect of the outer lensing profile on the
signal-to-noise ratio $\nu$ of weak lensing mass peaks. We have
considered two different filters, and found that $\nu$ can be biased
by $\la 4\%$ for peaks with $\nu\ga 3$ by adopting a single NFW
component in calculating $\nu$. Although this level of the bias is not
so large, it can be important in detailed comparisons of observed mass
peaks with theory in the future.

While in this paper we have focused on biases due to modeling
uncertainty of the outer lensing profiles, there are other effects
that could also bias the weak lensing analysis. Among others, the halo
triaxiality could have the most significant impact. Indeed, the large
triaxiality of cluster-scale haloes suggests that the lensing signal is
a strong function of the viewing angle \citep[e.g.,][]{oguri05}. 
We expect that the effect is almost averaged out by proper statistical
analysis \citep[e.g.,][]{corless09}, but the exact amount of the
residual bias due to the triaxiality should be sensitive to how the
cluster sample is selected.

We note that the analytic model considered in the paper is subject to
a potential improvement. For instance, the 2-halo term considered in
the paper is accurate in the large-scale limit. Near the cluster we
may have to consider various effects such as non-linearity and
stochasticity of the halo bias for more accurate theoretical
prediction. In addition, for more robust separation of signals between
the main halo and correlated matter, it is important to compare the
three-dimensional matter distribution around clusters with projected
lensing signals, which is left for future work. 

Nevertheless, our analytic models presented in the paper, which has
been tested against a large set of ray-tracing simulations, should be
useful for various aspects of cluster weak lensing studies as a handy
model of realistic lensing profiles in $N$-body simulations. In
particular, we expect that they are invaluable in the stacked weak
lensing study for which lensing signals can be detected much beyond
the virial radius \citep[e.g.,][]{menard10,umetsu11}. Given an
important cosmological information coming from the 2-halo term
\citep[e.g.,][]{oguri11}, it is essential to have a proper analytic
model which accurately predicted signals from the 1-halo to 2-halo
terms, in order for an unbiased interpretation of stacked lensing
data. The model is also useful for studying the effect of the outer
profile uncertainty (including the contribution from correlated matter
around clusters) on various cluster weak lensing analysis, as
explicitly demonstrated in the paper for shear profile fitting and
weak lensing mass peaks.  

\section*{Acknowledgments}

We thank M. Becker, S. Masaki, M. Sato, M. Takada, and N. Yoshida for
useful discussions.  This work is supported in part by Grant-in-Aid for
Scientific Research from the JSPS Promotion of Science (21740202).


\appendix

\section{Equations for lensing by the BMO profile}
\label{sec:bmo}

The lensing properties of the BMO profile (eq.~[\ref{eq:bmo}]) were
detailed in \citet{baltz09}, in which analytic expressions of lens
potential, deflection angle, convergence, etc. have been derived. Here
we reproduce key equations for both $n=1$ and $n=2$ cases with the
notation used in the paper.

For $n=1$, the convergence $\kappa(r)$ and average convergence
$\bar{\kappa}(<r)$ are given by
\begin{eqnarray}
\kappa(r)&=&\frac{4\rho_s r_s}{\Sigma_{\rm crit}}\frac{\tau^2}{2(\tau^2+1)^2}
\Biggl[\frac{\tau^2+1}{x^2-1}\left\{1-F(x)\right\}\nonumber\\
&&\hspace*{-8mm}+2F(x)-\frac{\pi}{\sqrt{\tau^2+x^2}}+\frac{\tau^2-1}{\tau\sqrt{\tau^2+x^2}}\,L(x)\Biggr],
\end{eqnarray}
\begin{eqnarray}
\bar{\kappa}(<r)&=&\frac{4\rho_s r_s}{\Sigma_{\rm crit}}
\frac{\tau^2}{(\tau^2+1)^2x^2}\Biggl[(\tau^2+2x^2+1)F(x)+\tau\pi\nonumber\\
&&\hspace*{-12mm}+(\tau^2-1)\ln\tau+
\sqrt{\tau^2+x^2}\left\{-\pi+\frac{\tau^2-1}{\tau}
L(x)\right\}\Biggr],
\end{eqnarray}
where $x\equiv r/r_s$, $\tau\equiv r_t/r_s=\tau_vc_{\rm vir}$, and
$F(x)$ and $L(x)$ are 
\begin{eqnarray}
 F(x)=\left\{ 
   \begin{array}{ll} 
     {\displaystyle \frac{1}{\sqrt{1-x^2}}{\rm arctanh}\sqrt{1-x^2}}&
     {\displaystyle (x<1),}\\
     {\displaystyle \frac{1}{\sqrt{x^2-1}}{\rm arctan}\sqrt{x^2-1}}&
     {\displaystyle (x>1),}
   \end{array} \right.
\end{eqnarray}
\begin{equation}
L(x)=\ln\left(\frac{x}{\sqrt{\tau^2+x^2}+\tau}\right).
\end{equation}
It is also useful to show the three-dimensional enclosed mass 
$m_{\rm bmo}(x)$ (see eq.[\ref{eq:mbmo_dim}]) for $n=1$. It is given by
\begin{eqnarray}
m_{\rm
  bmo}(x)&=&\frac{\tau^2}{2(\tau^2+1)^2(1+x)}\nonumber\\
&&\hspace*{-16mm}\times\Bigl[-2(\tau^2+1)x+4\tau(x+1){\rm arctan}(x/\tau)\nonumber\\
&&\hspace*{-16mm}+(\tau^2-1)(1+x)\ln(\tau^2(1+x)^2/(\tau^2+x^2))\Bigr].
\end{eqnarray}
Then the total mass $m_{\rm tot}$ (see eq.~[\ref{eq:mbmo_dim_tot}]) is
\begin{eqnarray}
m_{\rm tot}&\equiv & m_{\rm bmo}(\infty)
\nonumber\\
&=&\frac{\tau^2}{(\tau^2+1)^2}\left[\pi\tau-\tau^2-1+(\tau^2-1)\ln\tau\right].
\end{eqnarray}

For $n=2$, $\kappa(r)$ and $\bar{\kappa}(<r)$ are given by
\begin{eqnarray}
\kappa(r)&=&\frac{4\rho_s r_s}{\Sigma_{\rm crit}}\frac{\tau^4}{4(\tau^2+1)^3}
\Biggl[\frac{2(\tau^2+1)}{x^2-1}\left\{1-F(x)\right\}\nonumber\\
&&\hspace*{-8mm}+8F(x)
+\frac{\tau^4-1}{\tau^2(\tau^2+x^2)}-\frac{\pi[4(\tau^2+x^2)+\tau^2+1]}
{(\tau^2+x^2)^{3/2}}\nonumber\\
&&\hspace*{-8mm}+\frac{\tau^2(\tau^4-1)+(\tau^2+x^2)(3\tau^4-6\tau^2-1)}
{\tau^3(\tau^2+x^2)^{3/2}}\,L(x)\Biggr],
\end{eqnarray}
\begin{eqnarray}
\bar{\kappa}(<r)&=&\frac{4\rho_s r_s}{\Sigma_{\rm crit}}
\frac{\tau^4}{2(\tau^2+1)^3x^2}\Biggl[2(\tau^2+4x^2-3)F(x)\nonumber\\
&&\hspace*{-8mm}+\frac{1}{\tau}
\left\{\pi(3\tau^2-1)+2\tau(\tau^2-3)\ln \tau\right\}\nonumber\\
&&\hspace*{-8mm}+\frac{1}{\tau^3\sqrt{\tau^2+x^2}}\Bigl\{-\tau^3\pi
(4x^2+3\tau^2-1)+\nonumber\\
&&\hspace*{-8mm}+[2\tau^4(\tau^2-3)+x^2(3\tau^4-6\tau^2-1)]
L(x)\Bigr\}\Biggr].
\end{eqnarray}
The three-dimensional enclosed mass $m_{\rm bmo}(x)$ and the total mass
$m_{\rm tot}$ for $n=2$ were already given in equations
(\ref{eq:mbmo_nodim}) and (\ref{eq:mbmo_nodim_tot}), respectively.

From the above expressions, the tangential shear $\gamma_{\rm T}(r)$
and deflection angle $\phi_r(r)$ can be computed as
\begin{equation}
\gamma_{\rm T}(r)=\bar{\kappa}(<r)-\kappa(r),
\end{equation}
\begin{equation}
\phi_r(r)=r\,\bar{\kappa}(<r).
\end{equation}

\section{Fourier transform of the BMO profile}
\label{sec:fourier}

The Fourier transform of a dark halo density profile constitute an
important ingredient of the so-called halo model approach 
\citep[e.g.,][]{seljak00,ma00,scoccimarro01}. In this appendix we
study the Fourier transform of the BMO profile.

We consider the three-dimensional Fourier transform of
$u_M(r)\equiv\rho(r)/M_{\rm vir}$, $\tilde{u}_M(k)$ with $k$ being the
comoving wavenumber. We note that the Fourier transform of lensing
signals (convergence and tangential shear profiles) can also described
by using $\tilde{u}_M(k)$ as a consequence of the projection-slice
theorem \citep[see][]{oguri11}. We find that $\tilde{u}_M(k)$ for the BMO
profile can analytically be expressed as 
\begin{eqnarray}
\tilde{u}_M(k)&=&\frac{\tau}{2m_{\rm nfw}(c_{\rm vir})(1+\tau^2)^2x}\nonumber\\
&&\hspace*{-10mm}\times\Bigg[2(\tau^2-1)P(\tau x)-2\tau\pi e^{-\tau x}\nonumber\\
&&\hspace*{-6mm}+\tau\left\{\pi-2{\rm Si}(x)\right\}
\left\{2\cos x +(\tau^2+1)x\sin x\right\}
\nonumber\\
&&\hspace*{-6mm}+2\tau{\rm Ci}(x)
\left\{2\sin x -(\tau^2+1)x\cos x\right\}\Bigg],
\end{eqnarray}
for $n=1$, and 
\begin{eqnarray}
\tilde{u}_M(k)&=&\frac{\tau}{4m_{\rm nfw}(c_{\rm vir})(1+\tau^2)^3x}\nonumber\\
&&\hspace*{-10mm}\times\Bigg[2(3\tau^4-6\tau^2-1)P(\tau
x)-2\tau(\tau^4-1)xQ(\tau x)\nonumber\\
&&\hspace*{-6mm}-2\tau^2\pi e^{-\tau x}\left\{(\tau^2+1)x+4\tau\right\}\nonumber\\
&&\hspace*{-6mm}+2\tau^3\left\{\pi-2{\rm Si}(x)\right\}
\left\{4\cos x +(\tau^2+1)x\sin x\right\}
\nonumber\\
&&\hspace*{-6mm}+4\tau^3{\rm Ci}(x)
\left\{4\sin x -(\tau^2+1)x\cos x\right\}\Bigg],
\end{eqnarray}
for $n=2$, where $x\equiv (1+z)kr_s$. Note that $m_{\rm nfw}(c_{\rm
  vir})$ in the denominator originates from the division of the
density profile by $M_{\rm vir}$; if instead the profile is
normalised by the total mass $M_{\rm tot}$, $m_{\rm nfw}(c_{\rm
vir})$ should be replaced with $m_{\rm tot}$. The functions $P(x)$
and $Q(x)$ are defined as  
\begin{equation}
P(x)=\sinh(x){\rm Chi}(x)-\cosh(x){\rm Shi}(x),
\end{equation}
\begin{equation}
Q(x)=\frac{dP}{dx}=\cosh(x){\rm Chi}(x)-\sinh(x){\rm Shi}(x).
\end{equation}
For reference, the sine and cosine integrals and the hyperbolic sine
and cosine integrals are given by 
\begin{eqnarray}
{\rm Si}(x)&\equiv&\int_0^x\frac{\sin t}{t}dt,\\
{\rm Ci}(x)&\equiv&\gamma+\ln x+\int_0^x\frac{\cos t-1}{t}dt,\\
{\rm Shi}(x)&\equiv&\int_0^x\frac{\sinh t}{t}dt,\\
{\rm Chi}(x)&\equiv&\gamma+\ln x+\int_0^x\frac{\cosh t-1}{t}dt,
\end{eqnarray}
with $\gamma=0.57721566\ldots$ being the Euler-Mascheroni constant.
While $P(x)$ and $Q(x)$ can easily be computed numerically (except at
large $x$ where the round-off error dominates), we provide
fitting formulae for them:
\begin{eqnarray}
P_{\rm
  fit}(x)&=&-\left[\frac{1}{x}+\frac{bx^e}{c+(x-d)^2}\right]\left(\frac{x^4}{x^4+a^4}\right)^f\nonumber\\
&&\hspace*{-1mm}+x(\gamma+\ln x-1)\left(\frac{a^4}{x^4+a^4}\right)^f,
\end{eqnarray}
with ($a$, $b$, $c$, $d$, $e$, $f$)=($1.5652$, $3.38723$, $6.34891$,
$0.817677$, $-0.0895584$, $0.877375$), and 
\begin{eqnarray}
Q_{\rm fit}(x)&=&\left[\frac{1}{x^2}+\frac{bx^e}{c+(x-d)^4}\right]\left(\frac{x^4}{x^4+a^4}\right)^g\nonumber\\
&&\hspace*{-10mm}+\left[(\gamma+\ln x)\left(1+\frac{x^2}{2}\right)-\frac{3}{4}x^2\right]\left(\frac{a^4}{x^4+a^4}\right)^f,
\end{eqnarray}
with ($a$, $b$, $c$, $d$, $e$, $f$, $g$)=($2.26901$, $-2839.04$,
$265.511$, $-1.12459$, $-2.90136$, $1.86475$, $1.52197$). These
fitting formulae are obtained simply by connecting series expansions
at $x=0$ and $\infty$. We find that $P_{\rm fit}(x)$ is accurate at
better than 0.5\% for any positive values of $x$. The function 
$Q_{\rm fit}(x)$ is also accurate at better than 0.5\%, except at
around $x\approx0.88$ where $Q(x)$ has a root (i.e., $Q(x)\approx0$).

\label{lastpage}

\end{document}